\documentclass[12pt,a4paper,dvips]{article}
\usepackage{a4p}
\usepackage{cite,mcite,epsfig,rotating}
\usepackage{graphicx}
\usepackage{physics }
\usepackage{l3_title,ifthen, Lep}
\date{2 June 1999}
\lthreenote{XXXX}
\preprint{99-075}
\newcommand{\rar}{\rightarrow}

\def\mm{\mbox{$M \hskip 0.010in \! \! \! \! \! / _{ll}$}}

\newlength{\capindent}
\newlength{\capwidth}
\newlength{\figwidth}
\newcommand{\icaption}[2][!*!,!]{\hspace*{\capindent}%
  \begin{minipage}{\capwidth}
    \ifthenelse{\equal{#1}{!*!,!}}%
      {\caption{#2}}%
      {\caption[#1]{#2}}
  \end{minipage}}
\begin{document}
\bibliographystyle{l3style}
\begin{titlepage}
 \title{ Search for Heavy Neutral 
  and Charged Leptons \\ 
 in e$^+$e$^-$ Annihilation
 at {\boldmath $\sqrt{s} =$} 183 and 189 {\boldmath $\GeV$} }
\author{The L3 Collaboration}
%
%
\begin{abstract}
A search for unstable neutral and charged heavy leptons as well as for  
stable charged heavy leptons is performed at center-of-mass energies 
$\sqrt{s}$ = 183 and 189 \GeV\ 
with the L3 detector at LEP. No evidence for their 
existence is found. 
We exclude
neutral heavy leptons which couple to the electron, muon or tau family,
of the Dirac
type for masses below 92.4, 93.3  and 83.3 \GeV, and of the Majorana
type for masses below 81.8, 84.1 and 73.5 \GeV, respectively.
We exclude unstable charged heavy leptons for masses below
93.9 \GeV\ for a wide range of the associated neutral heavy lepton mass. 
If the unstable charged heavy lepton decays
to a light neutrino, we exclude masses below 92.4 \GeV.   
The production of stable charged heavy leptons with mass less than
93.5 \GeV\ is also excluded.
\end{abstract}

%
%
\submitted
 \end{titlepage}

\section*{Introduction}

Electron-positron colliders are well suited for
the search for new heavy leptons with masses up to
the beam energy~\cite{hlep01}.
Heavy leptons, L$^\pm$ or L$^0$, 
are pair-produced~\cite{hlep011} through the 
$s$-channel: e$^+$e$^- \rar \gamma/$Z$ \rar $L$^+$L$^-$, 
L$^0\overline{\mathrm{L}}^0$.
They are assumed to couple to the photon and the Z in the same way 
as the known leptons.
The total expected cross sections are in the range of 1 to 4 pb at masses well 
below the beam energy and fall as the mass of the lepton approaches 
the beam energy.
Here we report on a direct search for unstable sequential neutral 
heavy leptons (neutral partner to the charged lepton), 
L$^0$, of the Dirac or Majorana 
type, and charged heavy leptons, L$^{\pm}$, updating our previous search
~\cite{hlepl3}. 
Other recent results on this subject obtained at 
LEP at $\sqrt{s}$ = 133 to 183 \GeV, can be found in reference~\cite{hlep03}.
The data used in this analysis were collected with the L3 detector
at LEP at $\sqrt{s}$ = 183 \GeV\ with an integrated luminosity of 56 pb$^{-1}$,
and at $\sqrt{s}$ = 189 \GeV\ with an integrated 
luminosity of 176 pb$^{-1}$. We have combined the results with 
our earlier data recorded at  $\sqrt{s}$ = 133 to 172 \GeV.
The L3 detector is described elsewhere~\cite{hlep04}.

The neutral heavy lepton is expected to decay 
to a light lepton L$^0 \rar \ell^{\pm} $W$^{\mp*}$ ,
( $\ell =$ e, $\mu$, $\tau$). 
In this search we consider only the case where both neutral 
heavy leptons decay to the same lepton family (electron, muon, or tau).
In this case, the decay amplitude contains a parameter $U$ 
for the transition from the heavy lepton to the light lepton. 
The mean decay length, $D$, 
is given by~\cite{hlep05}
$D = \beta \gamma c \tau_{\mathrm{L}}  \propto \beta
 \vert U \vert^{-2} m_{\mathrm{L}}^{\alpha}$,
where $m_{\mathrm L}$ is the mass and $\tau_{\mathrm{L}}$ is the lifetime 
of the heavy lepton and $\alpha \approx -6$. 
To ensure high detection and reconstruction efficiencies, the search is
restricted to neutral leptons 
decaying within 1 cm of the interaction point. This
limits the sensitivity to the transition parameter to
$\vert U \vert {}^2$ $>$ $\mathcal{O}($10$^{-12})$.

Three different possibilities for the charged heavy lepton decay
modes are considered:

\begin{itemize}
\item[1)] The charged lepton decays via a lepton-number-nonconserving 
interaction to light neutrinos, L$^{\pm} \rar \nu_{\ell} $W$^{\pm*}$.
\item[2)] The charged lepton decays through a lepton-number-conserving
weak charged current interaction, L$^{\pm} \rar $L$^0 $W$^{\pm*}$,
with L$^0$ being stable.
\item[3)] The charged lepton is stable. This is the case 
         if the associated neutral lepton is heavier than its
         charged partner and there is no or very small probability to
         decay into light neutrinos. 
\end{itemize}

\section*{Event Simulation}

The generation of heavy leptons and their decay is performed with 
the TIPTOP~\cite{hlep06} Monte-Carlo program which 
takes into account initial state radiation and spin effects.
For the search we consider the 
mass range of the heavy leptons between 50 and 94 \GeV.
For the simulation of background from Standard Model processes the
following Monte-Carlo programs are used:
PYTHIA 5.7~\cite{hlep07} ($\mbox{e}^+ \mbox{e}^- \rar q \bar q (\gamma),
~\mbox{Z} \mbox{e}^+ \mbox{e}^-,~\mbox{Z} \mbox{Z}$),
KORALZ~\cite{hlep071} ($\mbox{e}^+ \mbox{e}^- \rar \tau^+ \tau^- (\gamma)$),
KORALW~\cite{hlep072} ($\mbox{e}^+ \mbox{e}^- \rar \mbox{W}^+ \mbox{W}^- $),
PHOJET~\cite{hlep08} ($\mbox{e}^+ \mbox{e}^- \rar \mbox{e}^+ \mbox{e}^- \mbox{q} \bar{\mbox{q}}$),
DIAG36~\cite{hlep081} ($\mbox{e}^+ \mbox{e}^- \rar \mbox{e}^+ \mbox{e}^- \tau^+ \tau^-$),
and EXCALIBUR~\cite{hlep082} ($\mbox{e}^+ \mbox{e}^- \rar \mbox{f} \bar{\mbox{f}}'
\mbox{f} \bar{\mbox{f}}'$). The Monte-Carlo events are 
simulated in the L3 detector using the GEANT3 program~\cite{hlep09}, 
which takes into account the effects of energy loss, 
multiple scattering and showering in the materials.

\section*{Search for Unstable Neutral Heavy Leptons}

The event topology used in the search for pair produced neutral heavy
leptons is two isolated leptons (e, $\mu$, or $\tau$) of the same 
family plus the decay products of real
or virtual W bosons,
i.e.  e$^+ $e$^- \rar $L$^0\overline{\mathrm{L}}^0
   \rar \ell^+\ell^- $W$^{+*}$W$^{-*}$ .
Hadronic events with visible energy greater than 60 \GeV\ and
charged track multiplicity greater than 3 are used in this analysis.
The search follows closely the procedure described 
previously~\cite{hlepl3}.

For the case where both neutral heavy leptons decay to either
electrons or muons, 
events must also satisfy the following criteria:

\begin{itemize}

\item The number of reconstructed jets plus isolated leptons is 
      at least 3.

\item The energy sum of the two isolated electrons or muons 
must be less than 70 \GeV.
This is a cut introduced to reject the Z-pair background.
Figure 1a) shows an example of the isolation criteria.
 
\end{itemize}

\noindent
After applying the selection, 6 events remain in 
the 189 \GeV\ data for the electron 
decay mode while 7.2$\pm$0.5 background events are expected.
For the muon decay mode, 1 event remains in the 189 \GeV\ data while
1.2$\pm$0.2 background events are expected. 
For the 183 \GeV\ data, 1 event satisfies the selection
requirements for the
electron decay mode and none for the muon decay mode
while 1.8$\pm$0.2 and 0.9$\pm$0.1  
background events are expected, respectively. 

For the case where both  neutral heavy leptons decay to tau leptons,
each of the taus can independently decay to hadrons, muons, or electrons.
When both tau leptons decay to either muons or electrons, the above 
selection is applied with the exception that we allow the isolated leptons 
to be either two electrons, two muons, or one muon and one electron.
We also consider the final state in which at least one of the 
tau decays into one charged hadron. 
In this case, events satisfying the 
following  criteria are selected: 

\begin{itemize}

\item The number of reconstructed jets plus isolated leptons 
          is at least 4.

\item The polar angle (angle with the beam axis) 
          $\theta$ of the missing momentum must be in the range
          $25^{\circ}<\theta<155^{\circ}$, and 
          the fraction of visible energy in the forward-backward region 
          ($\theta<20^{\circ}$ and $\theta>160^{\circ}$) must
          be less than 40\%.

\item The angle between the most isolated track and the track nearest to it,
          must be greater than 50$^{\circ}$ or the angle 
          between the second most isolated track and the track nearest to it,
          must be greater than 25$^{\circ}$.
          The transverse momenta, $p_t$, of the two most isolated tracks 
          must be greater than 1.2 \GeV, 
          and at least one track must have $p_t$ greater than 2.5 \GeV. 

\item The visible energy is required to be less than 175 \GeV.

\item The electron and muon energies must be less than 40 \GeV. 

\end{itemize}

\noindent
After applying the above selection, 33 events remain in the 189 \GeV\
data, while 32.3$\pm$1.0 background events are expected. For the 183 \GeV\
data, 5 events satisfy the selection requirements
while 6.8$\pm$0.3 background events are expected.  

The selection efficiencies are determined by Monte-Carlo. 
For neutral heavy leptons in the vicinity of the mass limit
from 80 to 94 \GeV, it is 28.0\% to 33.5\%
for the electron
decay mode.
For the muon decay mode the efficiency in the same mass range is 
25.0\% to 32.2\% .
For the tau decay mode, the selection
efficiency ranges from  23.4\% for  70 \GeV\ 
to 17.3\% for 85 \GeV. 
The systematic error,
which is mainly due to the uncertainties in the energy 
calibration factors and the lepton identification efficiency and
purity, is estimated to be 5\% relative. 
To obtain exclusion limits in this and
subsequent analyses, the selection efficiency has been reduced by one
standard deviation in the total systematic error. 
Taking into account the luminosity, 
the selection efficiency and the production
cross section we obtain a lower limit on the neutral heavy lepton mass
using the procedure from reference~\cite{hlep10}.
Combining this result with our previous analysis~\cite{hlepl3}
we exclude  at 95\% C.L.
the production of unstable neutral heavy leptons of the Dirac 
type for masses below 
92.4, 93.3 and 83.3 \GeV\ and of the Majorana type below 81.8, 84.1 
and 73.5 \GeV, if the neutral heavy lepton 
couples to the electron, muon and tau 
family, respectively. These results are summarized in Table~\ref{tab:Table1}.
         
\section*{Search for Unstable Charged Heavy Leptons }

\subsection*{Decay into a light neutrino, L$^{\pm} \rar \nu_{\ell} $W$^{\pm*}$}

For this search, two sets of cuts are used to search for the topology
with one hadronic and one leptonic W-decay as well as two hadronic
W-decays. For both selections we require that the visible energy be greater 
than 50 \GeV, the multiplicity of charged tracks be greater than 3, 
and the fraction of the total
visible energy in the forward-backward region 
be less than 25\%.

For the mode L$^+$L$^-$ $\rightarrow$ $\nu_{\ell} \bar\nu_{\ell}$
W$^{+*}$W$^{-*}$ 
$\rightarrow$ $\nu_{\ell} \bar\nu_{\ell} \ell \nu_{\ell} q \bar q'$,
events satisfying the following criteria are selected:

\begin{itemize}

\item The event contains at least one isolated electron or muon with
          energy greater than 4 \GeV\ and less than 50 \GeV.

\item The number of reconstructed jets plus isolated leptons is at least 3. 

\item The polar angle $\theta$ of the missing momentum must be in the range
          $25^{\circ}<\theta<155^{\circ}$.

\item     The sum of the energies of the hadronic jets must be
          less than 80 \GeV. Figure 1b) shows the distribution of the sum of 
          the energies of hadronic jets, after all other cuts have been 
          applied.

\end{itemize}

\noindent
After applying the selection, 
20 events remain in the 189 \GeV\ data while 23.2$\pm$0.8 background 
events are expected. To set mass limits, only the 189 \GeV\ data are
utilized since the lower energy data does not contribute significantly
to the setting of the limit.

For the mode where both W bosons decay hadronically, events satisfying 
the following criteria are selected:

\begin{itemize}

\item The event does not contain any isolated electrons or muons and
      the energy of each non-isolated electron or muon 
      must be less than 30 \GeV. 

\item The number of hadronic jets is at least 4.

\end{itemize}

For charged heavy lepton masses greater than the W boson mass, the leptons
each decay to a real W. 
The dominant remaining 
background after the previous cuts have been applied  
is W-pair production.
However, for the signal events, due to the energy and momentum carried
away by light neutrinos, the total visible energy is less than
$\sqrt{s}$ and the two W bosons are not back-to-back. 
To improve the determination of jet energies and angles
(both for the signal and the background) a kinematic fit is
applied imposing the constraint that both jet-jet invariant masses are
equal to the W mass. All possible jet-jet combinations are
considered and the one which gives the smallest $\chi^2$ of the fit is chosen.
The following additional requirements are then imposed for the case
where the charged lepton mass is greater than 80 \GeV:

\begin{itemize}

\item The visible energy must be less than 
      $\sqrt{s}-10$ \GeV. 
      Figure 1c)
      shows the distribution of visible energy, after all previous cuts 
      have been applied.

\item The angle between the two W candidates is less than 160$^{\circ}$.

\item The total transverse momentum of jets must be greater than 10 \GeV.
      Figure 1d) shows the distribution of this quantity, after all
      other cuts have been applied. 

\end{itemize}
\noindent
After applying the selection, 51 events remain in 189 \GeV\ data while 
53.3$\pm$1.5 background events are expected. To set mass limits only
the 189 \GeV\ data are utilized since the lower energy data does not 
contribute significantly to the setting of the limit.

The kinematic distributions of the candidates from both topologies 
are consistent with those expected from background.  
The selection efficiency for 90 \GeV\ charged heavy leptons is 
17.9\%. The systematic error, 
which is mainly due to the uncertainties in
the energy calibration factors and the lepton identification
efficiency and purity, 
is estimated to be 5\% relative.
Taking into account the luminosity, the selection efficiency 
and the production cross section 
we obtain a lower limit on the charged heavy lepton mass
using the procedure from reference~\cite{hlep10}.
We exclude the production of unstable charged heavy 
leptons at 95\% C.L. for masses below 92.4 \GeV. The results of this
search are summarized in Table~\ref{tab:Table1}.

\subsection*{Decay into stable neutral heavy lepton, 
L$^{\pm} \rar $L$^0 $W$^{\pm*}$}

In this case the charged lepton decays into its associated neutral
lepton L$^0$.
From LEP results at the Z resonance~\cite{hlep11}, the mass of the 
stable L$^0$ must be greater than 40 \GeV, thus the signal events are 
characterized by a large missing energy and a large 
transverse momentum imbalance. 
In the limit of a vanishing mass
difference between charged lepton and associated neutral lepton
($\Delta m$ = $m_{\rm{L}^{\pm}}-m_{\rm L^0}$), 
the sensitivity to the signal
is limited by the trigger efficiency and the large  
two-photon background. 
Hence, the search is restricted to 
10 \GeV\ $\leq$ $\Delta m$ $\leq$ 45 \GeV.
The case of a light neutral lepton ($\Delta m$ = $m_{\rm L^{\pm}}$) 
has been considered in the previous section.
The main background is the two-photon process for small mass
difference ($\Delta m~\leq$ 20 \GeV) and the 
$q \bar q (\gamma)$ and WW processes
for high mass difference ($\Delta m~\geq$ 20 \GeV).

The above signature of a charged heavy lepton is very similar
to that of a chargino, when the chargino decays into a stable
neutralino and a W boson. Therefore, we use
a selection developed for the chargino search~\cite{hlep12},
which is mainly based on the signatures of missing energy,
transverse momentum imbalance,
missing mass, acoplanarity and isolated leptons. 

After applying the selection, 9 events remain in the 189 \GeV\
data while 9.9$\pm$1.5  events are expected from background.
Only the 189 \GeV\ data are utilized since the lower energy
data does not contribute significantly to the setting of the limit.

The selection efficiency varies from $\sim$5\% for $\Delta m$ = 10 \GeV\
to 40\% for $\Delta m$ = 45 \GeV.
The typical systematic error, which is mainly
due to the the Monte-Carlo statistics
and to uncertainties in energy
calibration factors and jet angular resolution, is estimated to be
5\% relative.
Taking into account the luminosity, selection efficiency and the production
cross section we obtain a lower limit on the charged heavy
lepton mass. Figure 2 shows the 95\% C.L. exclusion contour in the
$m_{\mathrm{L^{\pm}}}-m_{\mathrm{L^0}}$ plane. The exclusion 
region extends to  $m_{\mathrm{L^{\pm}}}=$ 93.9 \GeV. The result of this
search is summarized in Table~\ref{tab:Table1}.

\section*{Search for Stable Charged Heavy Leptons}

This search is designed for stable charged particles with masses from 45 \GeV\ up to 
the beam energy, $E_B$. 
The search uses all the data from  $\sqrt{s}$ = 133 \GeV\ to 189 \GeV, 
and follows closely the procedure described in
reference~\cite{hlep14} for the
stable slepton-pair search.
The main difference between the two analyses is that the efficiency 
is greater for sleptons which
have a different angular distribution than fermions.
The search is performed with selection requirements
optimized for three different mass regions: high ($m_{\rm L}/E_B>$ 0.8),
intermediate (0.7 $<m_{\rm L}/E_B<$ 0.8)
and low (0.5 $<m_{\rm L}/E_B<$ 0.7).

Events are selected which have two charged tracks with momentum
greater than 5 \GeV\ and polar angle $\rm\vert \cos\theta\vert
< 0.82$. The acollinearity angle between the two tracks is required to be
less than $\rm 15^o$.
The polar angle requirement selects events where the
trigger and track reconstruction efficiency is high and the
$dE/dx$ resolution in the tracking chamber is good. 
The momentum and acollinearity angle cuts reduce the background from two photon
produced lepton pairs as well as from  
dilepton annihilation events with a high energy photon in the
final state.  

In the region $m_{\rm L}/E_B>$ 0.8, the stable leptons are highly ionizing and
only the tracking chamber information is utilized. 
The $dE/dx$ measurement is calibrated with
Bhabha scattering events, and the mean value of the resulting $dE/dx$ distribution is
normalized to one. Its resolution is 0.08.
Events are selected for which
the ionization energy loss for each track
is between  1.25 and 8, and the product of the track ionization losses 
is larger than 2.

Figure 3 shows the $dE/dx$ measurements 
of the first and second tracks for data events 
at 189 \GeV\ passing all cuts except those on track ionization,
as well as the simulated signal from pair-production of 93 GeV
stable charged leptons.
One candidate event satisfies the selection requirements 
in the 189 \GeV\ data and none in the lower energy data
samples.  
The candidate event in the data corresponds to pair production of stable
charged particles of mass $80^{+1.5}_{-2.5}$ \GeV. 
The background is estimated to be less than 0.5 events at the 90\% C.L.
for the entire data sample. We have conservatively assumed that 
there are no background
events when producing exclusion limits in this region.

The efficiency for selection of heavy stable charged leptons ranges from
58\% to 70\% over the range  0.8 $<m_{\rm L}/E_B<$ 0.99. 
For $m_{\rm L}/E_B>$ 0.99, 
the charged particles are very highly ionizing and saturation
effects become significant and may lead to track reconstruction inefficiencies.
The ionization has been studied using low energy protons in our hadronic
events up to 8 units in $dE/dx$ which corresponds to $m_{\rm L}/E_B=$ 0.99.


In the region 0.7 $< m_{\rm L}/E_B<$ 0.8, we combine the information from
the $dE/dx$ and the muon chambers to search for stable heavy leptons.
Events are required to satisfy the following criteria:
the matched energy deposit in the electromagnetic calorimeter 
is less than 2 \GeV\ for each track; the ionization energy loss 
for each track is greater than 1.05,  their product is larger 
than 1.25, and there is at least one track in the muon chamber
with $p/E_B\rm >0.4$.

One candidate event is observed in the 189 \GeV\ data sample and 
none in the lower energy data samples. From the $dE/dx$ information,
the candidate event is consistent with pair production of 
mass 64$\pm$ 3 \GeV\ stable charged particles.
The total background in the entire data sample
is estimated to be 1.1$\pm$0.2 events 
of which 0.6 events are expected in the 189 \GeV\ data.
This background is taken into account
in producing exclusion limits.
The selection efficiency ranges
from 34\% at $m_{\rm L}/E_B=$ 0.7 to 58\% at $m_{\rm L}/E_B=$ 0.8. 


In the region 0.5 $<m_{\rm L}/E_B<$ 0.7, 
the $dE/dx$ ionization is indistinguishable from that of lighter particles.
Consequently, only the muon chamber momentum and calorimeter
information are used. Assuming
the energy of the particles is $E_B$, the momentum of the 
track can be used to reconstruct the mass of the heavy particle:
$m_{\rm L} = \sqrt{{E_B}^2 - p^2}$.
Events are selected which satisfy the following criteria:
two tracks are found in the muon spectrometer in 
the polar angle range $\vert \cos\theta \vert<0.76$,
each with 0.55 $< p/E_B<$ 1.0;
the acollinearity angle between the tracks is less than $10^{\circ}$;
the matched energy deposit in the electromagnetic calorimeter  
is less than 2 \GeV\ and that in the hadronic calorimeter less than 
15 \GeV\ for each track;
the sum of the unmatched electromagnetic calorimeter 
energy deposits is less than 1 \GeV;
the reconstructed mass using the muon chamber momentum for each particle
exceeds 45 \GeV.

Three events pass the selection requirements in the 189 \GeV\ data 
and none in the lower energy data samples.
The background is estimated
to be 3.3$\pm$0.1 events in the entire data sample
of which 2.6 events are expected in the 189 \GeV\ data, mostly from
dimuon events. The background is taken into account in determining
exclusion limits from the data.
The selection efficiency
ranges from 9\% at $m_{\rm L}/E_B =$ 0.5 to 34\% at $m_{\rm L}/E_B=$ 0.7.


The upper limit on the number of signal events 
over the entire mass range 45 \GeV\ to 93.5 \GeV\ has been converted
to an upper limit on the production cross section using the luminosity 
of the data and the selection efficiency. 
The systematic error, estimated to be 5\% relative, is 
mainly due to the Monte-Carlo statistics.  
%
%

Figure 4 shows the upper limit of the cross section obtained from
the combined 133 \GeV\ to 189 \GeV\ data. 
The candidate events are accounted for by using a range of $\pm 2\sigma$ 
in the uncertainty in the mass centered about the mass 
corresponding to the observed events.
The 95\% CL. upper limit on the 
production cross section for pair production of stable heavy leptons
is  0.08-0.02 pb for the mass range 50 to 93.5 \GeV\
and less than 0.18 pb for the mass range 45 to 50 \GeV.
We do not quote an upper limit for masses greater than 93.5 \GeV.

Figure 4 also shows the calculated cross section~\cite{hlep06}
for heavy charged leptons as a function of mass
at $\rm \sqrt{s}=$189 \GeV.
From the comparison of the two curves,
we exclude production of stable heavy charged leptons
with mass less than 93.5 \GeV\ at 95\% C.L. and this result is included in
Table~\ref{tab:Table1}.

\section*{Acknowledgements}

We wish to express our gratitude to the CERN accelerator divisions for the 
excellent performance of the LEP machine. We acknowledge with appreciation 
the effort of the engineers, technicians and support staff who have
participated in the construction and maintenance of this experiment.

\begin{table}[tb!]
\caption{Mass limits from the various heavy lepton search channels.}
\label{tab:Table1} 
\vspace*{0.2cm}
\begin{center}
{\begin{tabular}{|c|cc|} \hline
 Channel & \multicolumn{2}{c|}{\hspace*{.5cm}
95\% CL Mass Limit \GeV~~} \\\hline
$\rm L^0$        & \hspace*{.5cm} Dirac & \hspace*{.5cm} Majorana \\ 
$\rm L^0 \rightarrow eW $       &  \hspace*{.5cm}  92.4    & 
 \hspace*{.5cm} 81.8  \\
$\rm L^0 \rightarrow \mu W$     &  \hspace*{.5cm}  93.3    &  
 \hspace*{.5cm} 84.1  \\
$\rm L^0 \rightarrow \tau W$    &  \hspace*{.5cm}  83.3    &  
 \hspace*{.5cm} 73.5  \\ \hline \hspace*{.5cm}
$\rm L^\pm \rightarrow \nu_\ell W^{\pm *}$ & 
\multicolumn{2}{c|} {92.4} \\ \hline
$\rm L^\pm \rightarrow L^0 W^{\pm *}$ & 
\multicolumn{2}{c|} {93.9} \\ \hline
Stable $\rm L^\pm$ & \multicolumn{2}{c|}{93.5} \\ \hline
\end{tabular}} 
\end{center}
\end{table}

%
%
\newpage
\section*{Author List}
\typeout{   }     
\typeout{Using author list for paper 175 -?}
\typeout{$Modified: Tue May 18 08:38:13 1999 by clare $}
\typeout{!!!!  This should only be used with document option a4p!!!!}
\typeout{   }
%
%
%
%
%
%

\newcount\tutecount  \tutecount=0
\def\tutenum#1{\global\advance\tutecount by 1 \xdef#1{\the\tutecount}}
\def\tute#1{$^{#1}$}
\tutenum\aachen            
\tutenum\nikhef            
\tutenum\mich              
\tutenum\lapp              
\tutenum\basel             
\tutenum\lsu               
\tutenum\beijing           
\tutenum\berlin            
\tutenum\bologna           
\tutenum\tata              
\tutenum\ne                
\tutenum\bucharest         
\tutenum\budapest          
\tutenum\mit               
\tutenum\florence          
\tutenum\cern              
\tutenum\wl                
\tutenum\geneva            
\tutenum\hefei             
\tutenum\seft              
\tutenum\lausanne          
\tutenum\lecce             
\tutenum\lyon              
\tutenum\madrid            
\tutenum\milan             
\tutenum\moscow            
\tutenum\naples            
\tutenum\cyprus            
\tutenum\nymegen           
\tutenum\caltech           
\tutenum\perugia           
\tutenum\cmu               
\tutenum\prince            
\tutenum\rome              
\tutenum\peters            
\tutenum\salerno           
\tutenum\ucsd              
\tutenum\santiago          
\tutenum\sofia             
\tutenum\korea             
\tutenum\alabama           
\tutenum\utrecht           
\tutenum\purdue            
\tutenum\psinst            
\tutenum\zeuthen           
\tutenum\eth               
\tutenum\hamburg           
\tutenum\taiwan            
\tutenum\tsinghua          
{
\parskip=0pt
\noindent
{\bf The L3 Collaboration:}
\ifx\selectfont\undefined
 \baselineskip=10.8pt
 \baselineskip\baselinestretch\baselineskip
 \normalbaselineskip\baselineskip
 \ixpt
\else
 \fontsize{9}{10.8pt}\selectfont
\fi
\medskip
\tolerance=10000
\hbadness=5000
\raggedright
\hsize=162truemm\hoffset=0mm
\def\r{\rlap,}
\noindent

M.Acciarri\r\tute\milan\
P.Achard\r\tute\geneva\ 
O.Adriani\r\tute{\florence}\ 
M.Aguilar-Benitez\r\tute\madrid\ 
J.Alcaraz\r\tute\madrid\ 
G.Alemanni\r\tute\lausanne\
J.Allaby\r\tute\cern\
A.Aloisio\r\tute\naples\ 
M.G.Alviggi\r\tute\naples\
G.Ambrosi\r\tute\geneva\
% ...
\newpage
\begin{list}{A}{\itemsep=0pt plus 0pt minus 0pt\parsep=0pt plus 0pt minus 0pt
                \topsep=0pt plus 0pt minus 0pt}
\item[\aachen]
 I. Physikalisches Institut, RWTH, D-52056 Aachen, FRG$^{\S}$\\
 III. Physikalisches Institut, RWTH, D-52056 Aachen, FRG$^{\S}$
\item[\nikhef] National Institute for High Energy Physics, NIKHEF, 
     and University of Amsterdam, NL-1009 DB Amsterdam, The Netherlands
\item[\mich] University of Michigan, Ann Arbor, MI 48109, USA
\item[\lapp] Laboratoire d'Annecy-le-Vieux de Physique des Particules, 
     LAPP,IN2P3-CNRS, BP 110, F-74941 Annecy-le-Vieux CEDEX, France
\item[\basel] Institute of Physics, University of Basel, CH-4056 Basel,
     Switzerland
\item[\lsu] Louisiana State University, Baton Rouge, LA 70803, USA
\item[\beijing] Institute of High Energy Physics, IHEP, 
  100039 Beijing, China$^{\triangle}$ 
\item[\berlin] Humboldt University, D-10099 Berlin, FRG$^{\S}$
\item[\bologna] University of Bologna and INFN-Sezione di Bologna, 
     I-40126 Bologna, Italy
\item[\tata] Tata Institute of Fundamental Research, Bombay 400 005, India
\item[\ne] Northeastern University, Boston, MA 02115, USA
\item[\bucharest] Institute of Atomic Physics and University of Bucharest,
     R-76900 Bucharest, Romania
\item[\budapest] Central Research Institute for Physics of the 
     Hungarian Academy of Sciences, H-1525 Budapest 114, Hungary$^{\ddag}$
\item[\mit] Massachusetts Institute of Technology, Cambridge, MA 02139, USA
\item[\florence] INFN Sezione di Firenze and University of Florence, 
     I-50125 Florence, Italy
\item[\cern] European Laboratory for Particle Physics, CERN, 
     CH-1211 Geneva 23, Switzerland
\item[\wl] World Laboratory, FBLJA  Project, CH-1211 Geneva 23, Switzerland
\item[\geneva] University of Geneva, CH-1211 Geneva 4, Switzerland
\item[\hefei] Chinese University of Science and Technology, USTC,
      Hefei, Anhui 230 029, China$^{\triangle}$
\item[\seft] SEFT, Research Institute for High Energy Physics, P.O. Box 9,
      SF-00014 Helsinki, Finland
\item[\lausanne] University of Lausanne, CH-1015 Lausanne, Switzerland
\item[\lecce] INFN-Sezione di Lecce and Universit\'a Degli Studi di Lecce,
     I-73100 Lecce, Italy
\item[\lyon] Institut de Physique Nucl\'eaire de Lyon, 
     IN2P3-CNRS,Universit\'e Claude Bernard, 
     F-69622 Villeurbanne, France
\item[\madrid] Centro de Investigaciones Energ{\'e}ticas, 
     Medioambientales y Tecnolog{\'\i}cas, CIEMAT, E-28040 Madrid,
     Spain${\flat}$ 
\item[\milan] INFN-Sezione di Milano, I-20133 Milan, Italy
\item[\moscow] Institute of Theoretical and Experimental Physics, ITEP, 
     Moscow, Russia
\item[\naples] INFN-Sezione di Napoli and University of Naples, 
     I-80125 Naples, Italy
\item[\cyprus] Department of Natural Sciences, University of Cyprus,
     Nicosia, Cyprus
\item[\nymegen] University of Nijmegen and NIKHEF, 
     NL-6525 ED Nijmegen, The Netherlands
\item[\caltech] California Institute of Technology, Pasadena, CA 91125, USA
\item[\perugia] INFN-Sezione di Perugia and Universit\'a Degli 
     Studi di Perugia, I-06100 Perugia, Italy   
\item[\cmu] Carnegie Mellon University, Pittsburgh, PA 15213, USA
\item[\prince] Princeton University, Princeton, NJ 08544, USA
\item[\rome] INFN-Sezione di Roma and University of Rome, ``La Sapienza",
     I-00185 Rome, Italy
\item[\peters] Nuclear Physics Institute, St. Petersburg, Russia
\item[\salerno] University and INFN, Salerno, I-84100 Salerno, Italy
\item[\ucsd] University of California, San Diego, CA 92093, USA
\item[\santiago] Dept. de Fisica de Particulas Elementales, Univ. de Santiago,
     E-15706 Santiago de Compostela, Spain
\item[\sofia] Bulgarian Academy of Sciences, Central Lab.~of 
     Mechatronics and Instrumentation, BU-1113 Sofia, Bulgaria
\item[\korea] Center for High Energy Physics, Adv.~Inst.~of Sciences
     and Technology, 305-701 Taejon,~Republic~of~{Korea}
\item[\alabama] University of Alabama, Tuscaloosa, AL 35486, USA
\item[\utrecht] Utrecht University and NIKHEF, NL-3584 CB Utrecht, 
     The Netherlands
\item[\purdue] Purdue University, West Lafayette, IN 47907, USA
\item[\psinst] Paul Scherrer Institut, PSI, CH-5232 Villigen, Switzerland
\item[\zeuthen] DESY-Institut f\"ur Hochenergiephysik, D-15738 Zeuthen, 
     FRG
\item[\eth] Eidgen\"ossische Technische Hochschule, ETH Z\"urich,
     CH-8093 Z\"urich, Switzerland
\item[\hamburg] University of Hamburg, D-22761 Hamburg, FRG
\item[\taiwan] National Central University, Chung-Li, Taiwan, China
\item[\tsinghua] Department of Physics, National Tsing Hua University,
      Taiwan, China
\item[\S]  Supported by the German Bundesministerium 
        f\"ur Bildung, Wissenschaft, Forschung und Technologie
\item[\ddag] Supported by the Hungarian OTKA fund under contract
numbers T019181, F023259 and T024011.
\item[\P] Also supported by the Hungarian OTKA fund under contract
  numbers T22238 and T026178.
\item[$\flat$] Supported also by the Comisi\'on Interministerial de Ciencia y 
        Tecnolog{\'\i}a.
\item[$\sharp$] Also supported by CONICET and Universidad Nacional de La Plata,
        CC 67, 1900 La Plata, Argentina.
\item[$\natural$] Supported by Deutscher Akademischer Austauschdienst.
\item[$\diamondsuit$] Also supported by Panjab University, Chandigarh-160014, 
        India.
\item[$\triangle$] Supported by the National Natural Science
  Foundation of China.
\end{list}
}
\vfill




\newpage




\newpage


\begin{figure}[p]
\begin{center}
\mbox{\epsfysize=16cm\epsffile{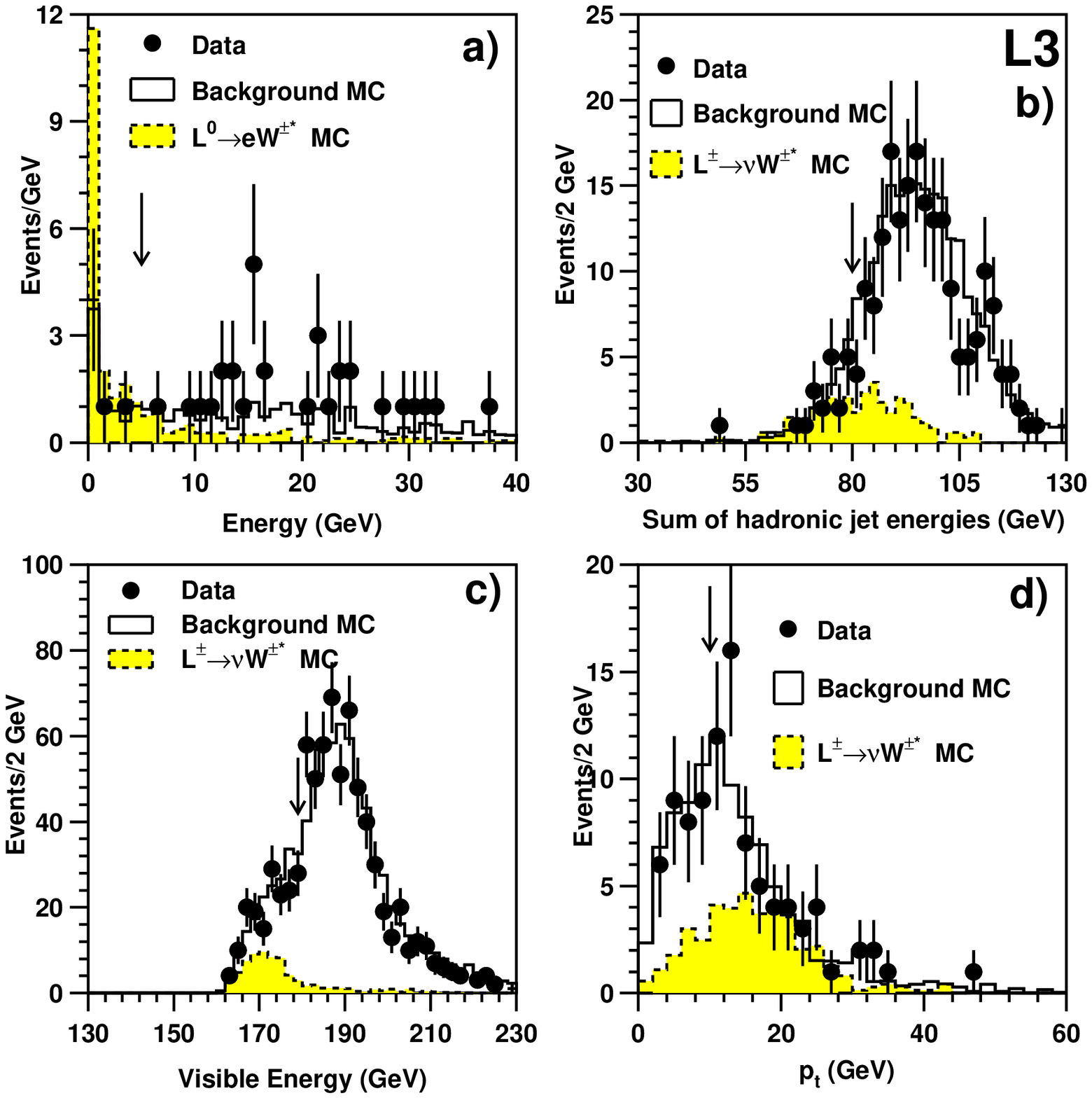}}
\end{center}
\caption{
a) Energy in a 30$^{\circ}$ cone around the second most energetic electron
candidate.
b) The sum of the energies of the hadronic jets.
c) Visible energy in the event.
d) The distribution of the total transverse momentum, $p_t$.
The dots are the data, the solid histogram
is the background Monte-Carlo. The dashed line represents the simulated
signal for e$^+ $e$^- \rar \mathrm{L}\overline{\mathrm{L}}$ from the 
TIPTOP Monte-Carlo.
The normalization for the signal Monte-Carlo is scaled by a factor of 2
for better visibility. 
The arrows indicate the corresponding values of the applied cuts.
In this case, all cuts on other quantities have been applied.
}
\label {4fig}
\end{figure}
\begin{figure}[p]
\begin{center}
\mbox{\epsfysize=16cm\epsffile{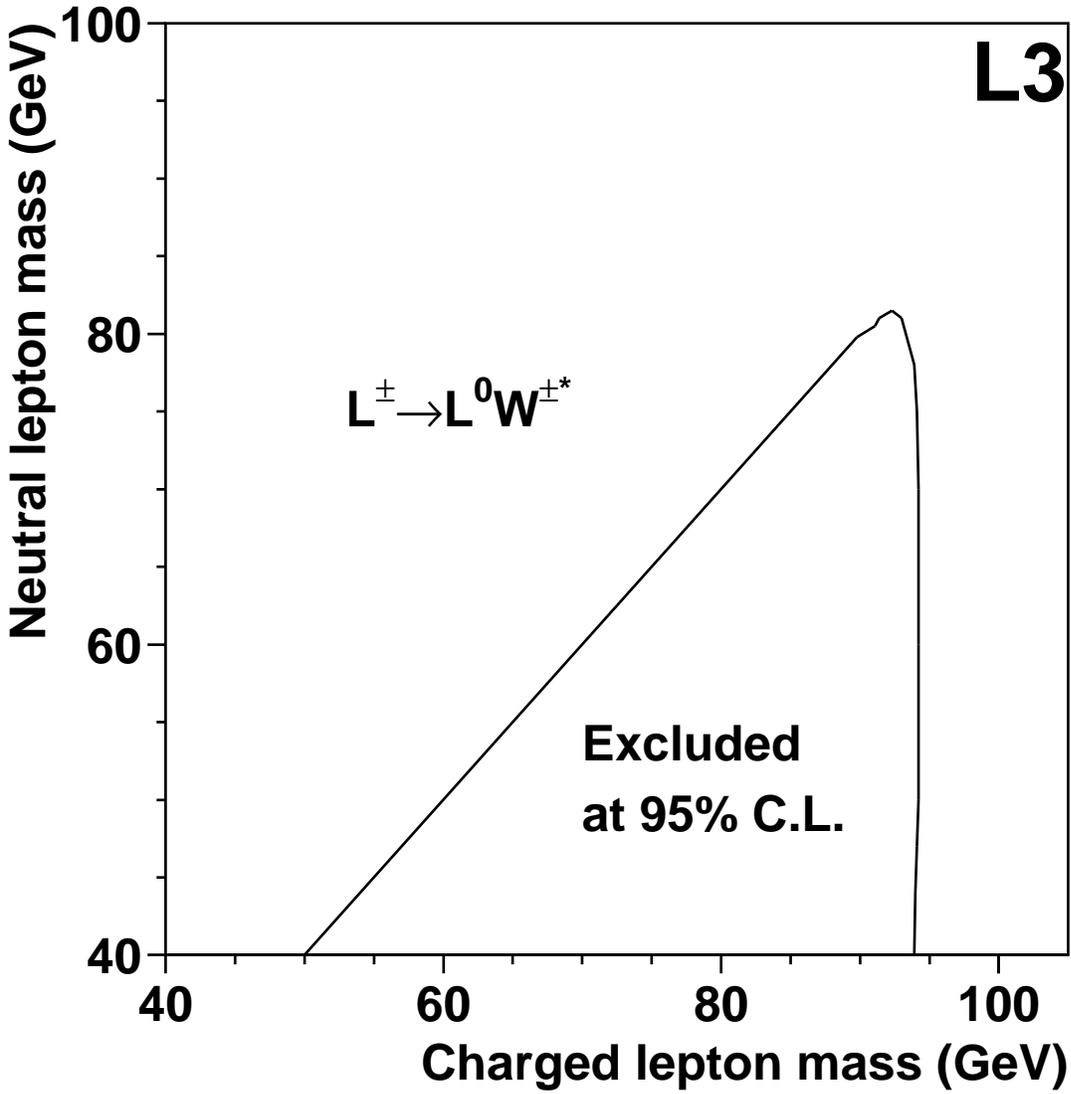}}
\end{center}
\caption{
The 95\% confidence level limit on the charged heavy
lepton mass $m_{\mathrm{L^{\pm}}}$ and the associated neutral heavy
lepton mass $m_{\mathrm{L^0}}$ assuming that the L$^0$ is stable. 
}
\label {clcn}
\end{figure}

\begin{figure}[p]
\begin{center}
\mbox{\epsfysize=16cm\epsffile{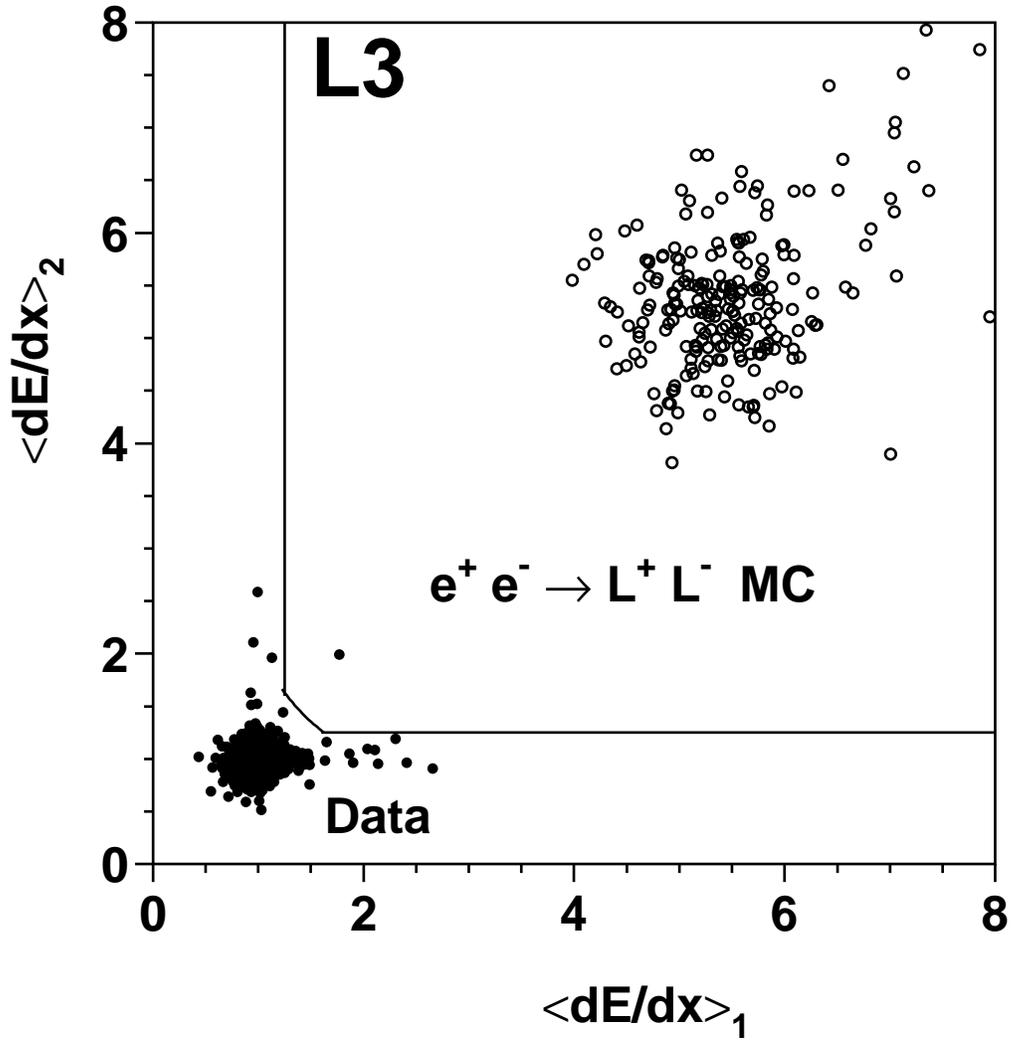}}
\end{center}
\caption{
The normalized measured track energy loss of track 2 
vs track 1 for the data (solid circles) taken at 
$\sqrt{s}$ = 189 \GeV\ 
and the simulated signal (arbitrary nomalization) for a mass
93 \GeV\ stable heavy lepton (open circles). The lines represent
the applied cut. 
}
\label {fig06}
\end{figure}

\begin{figure}[p]
\begin{center}
\mbox{\epsfysize=16cm\epsffile{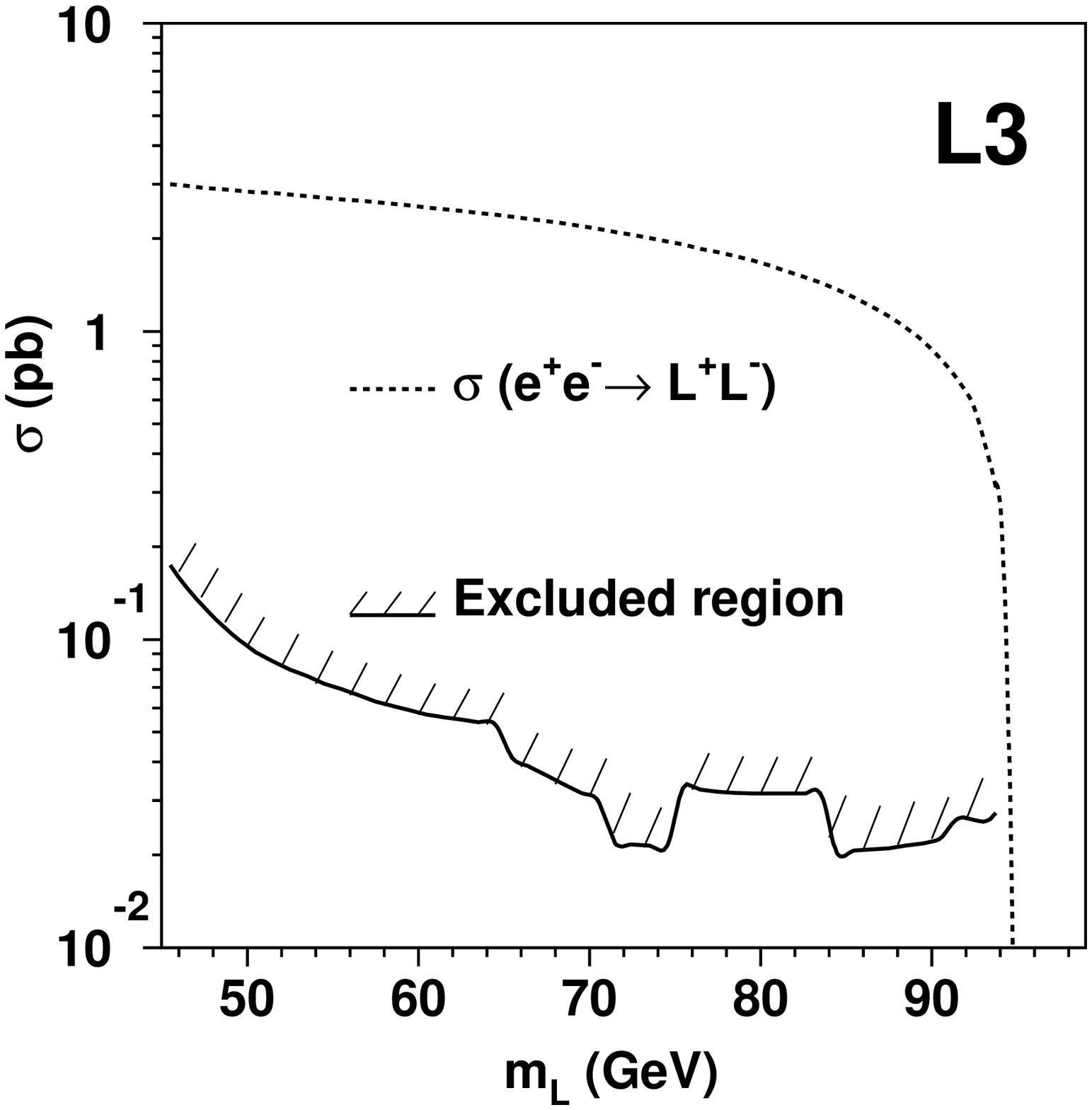}}
\end{center}
\caption{
Upper limit of the production cross section for 
pair-production of stable charged heavy leptons
in the L3 data at $\sqrt{s}$ = 133 to 189 \GeV\ as a function of mass.
We do not determine the upper limit for masses greater than 93.5 \GeV.
The hatched area indicates the region excluded by our search.
The dashed line represents the calculated pair-production cross section for 
heavy leptons at $\sqrt{s}$ = 189 \GeV\ from the TIPTOP Monte-Carlo.
}
\label {fig07}
\end{figure}


\begin{thebibliography}{10}                                                     

\bibitem{hlep01}
A. Zichichi {\em et al.}, Preprint INFN/AE-67/3; Lett. Nuovo Cimento {\bf 4} 
(1970) 1156; Nuovo Cimento {\bf 17 A} (1973) 383;\newline
M. Perl {\em et al.}, Phys. Rev. Lett. {\bf 35} (1975) 1489.

\bibitem{hlep011}
See for example the review by A.~Djouadi, J.~Ng and T.G.~Rizzo, 
SLAC-PUB-95-6772, in {\it Electroweak Symmetry Breaking and
New Physics at the TeV Scale}, T.~Barklow, S.~Dawson, H.E.~Habor and 
S.~Siegrist, Singapore, World Scientific (1997).

\bibitem{hlepl3}L3 Collaboration, M. Acciarri {\em et al.}, 
Phys. Lett. {\bf B 377} (1996) 304;\newline
L3 Collaboration, M. Acciarri {\em et al.}, 
Phys. Lett. {\bf B 412} (1997) 189.


\bibitem{hlep03}
ALEPH Collaboration, D. Buskulic {\em et al.}, Phys. Lett. {\bf B 384} (1996) 
439;\newline
ALEPH Collaboration, R. Barate {\em et al.}, Phys. Lett. {\bf B 405} (1997)
379;\newline
DELPHI Collaboration, P. Abreu {\em et al.}, Phys. Lett. {\bf B 396} (1997)
315;\newline
DELPHI Collaboration, P. Abreu {\em et al.}, 
E. Phys. J. {\bf C8} (1999) 41;\newline
DELPHI Collaboration, P. Abreu {\em et al.}, 
Phys. Lett. {\bf B444} (1998) 491;\newline
OPAL Collaboration, G. Alexander {\em et al.}, Phys. Lett. {\bf B 385} (1996)
433;\newline
OPAL Collaboration, K. Ackerstaff {\em et al.}, Phys. Lett. {\bf B 393} (1997)
217.\newline
OPAL Collaboration, K. Ackerstaff {\em et al.}, Eur. Phys. J. {\bf C 1} (1998)
45.



\bibitem{hlep04}                                                           
L3 Collaboration, B. Adeva {\em et al.}, Nucl. Instr. Meth. {\bf A 289}
(1990) 35;\newline
M. Acciari {\em et al.}, Nucl. Instr. Meth. {\bf A 351}
(1994) 300;\newline 
M. Chemarin {\em et al.}, Nucl. Instr. Meth. {\bf A 349} (1994) 345;\newline
M. Adam {\em et al.}, Nucl. Instr. Meth. {\bf A 383} (1996) 342;\newline
G. Basti {\em et al.}, Nucl. Instr. Meth. {\bf A 374} (1996) 293.

\bibitem{hlep05}
M. Gronau, C. Leung and J. Rosner, Phys. Rev. {\bf D29} (1984) 2539.

\bibitem{hlep06}
S. Jadach and J. K\"uhn, 
Preprint MPI-PAE/PTh 64/86.

\bibitem{hlep07} 
T.~Sj\"ostrand, Comp. Phys. Comm. {\bf 82} (1994) 74.

\bibitem{hlep071}
S. Jadach, B.F.L.Ward and Z. Was, Comp. Phys. Comm. {\bf 79} (1994) 503.

\bibitem{hlep072} 
M. Skrzypek {\em et al.}, Comp. Phys. Comm. {\bf 94} (1996) 216;\\
M. Skrzypek {\em et al.}, Phys. Lett. {\bf B 372} (1996) 289.

\bibitem{hlep08}
R. Engel, Z. Phys. {\bf C 66} (1993) 1657;\\
R. Engel and J. Ranft, Phys. Rev. {\bf D 54} (1996) 4244.

\bibitem{hlep081}
F.A. Berends, P.H. Daverveldt and R. Kleiss, Nucl. Phys. {\bf 
B 253} (1985) 421; Comp. Phys. Comm. {\bf 40} (1986) 271.

\bibitem{hlep082}
F.A. Berends, R. Kleiss and R. Pittau, Nucl. Phys. {\bf B424} (1994) 308;
Nucl. Phys. {\bf B426} (1994) 344; Nucl. Phys. (Proc. Suppl.) {\bf B37} 
(1994) 163; Phys. Lett. {\bf B 335} (1994) 490; \\
R. Kleiss and R. Pittau, Comp. Phys. Comm. {\bf 83} (1994) 141.

\bibitem{hlep09}
The L3 detector simulation is based on GEANT Version 3.15. See R. Brun et al.,
"GEANT 3", CERN DD/EE/84-1 (Revised), September 1987. The GHEISHA program (H.
Fesefeldt, RWTH Aachen Report PITHA 85/02 (1985)) is used to simulate hadronic
interactions.


\bibitem{hlep10}
V.F.~Obraztsov, Nucl. Instr. Meth. {\bf A 316} (1992) 388.

\bibitem{hlep11}
ALEPH Collaboration, D. Decamp {\em et al.}, Phys. Lett. {\bf B 236} (1990) 
511;\newline
DELPHI Collaboration, P. Abreu {\em et al.}, Phys. Lett. {\bf B 274} (1992)
230;\newline
L3 Collaboration, B.~Adeva  {\em et al},  Phys. Lett. {\bf B 251} (1990) 
321;\newline
OPAL Collaboration, G. Alexander {\em et al.}, Z. Phys. {\bf C 52} (1991)
200.



\bibitem{hlep12}
L3 Collaboration, M. Acciarri {\em et al.}, Eur. Phys. Journal {\bf C 4},
(1998) 207.

\bibitem{hlep14}
L3 Collab., M.~Acciarri {\it et al.}, CERN EP/99-024,
to be published in Phys Lett {\bf B}.

\end{thebibliography}
\end{document}